\documentclass[12pt]{article}

\usepackage{color}
\usepackage{epsfig,
amssymb,
amsfonts,
amsmath,
graphicx,
dsfont,
xfrac,
authblk,
subcaption
}

\definecolor{mygray}{gray}{0.5}

\usepackage[normalem]{ulem}

\usepackage{cite}
\usepackage[colorlinks=true,linkcolor=blue,citecolor=red]{hyperref}

\newcommand{\be}{\begin{equation}}
\newcommand{\ee}{\end{equation}}
\newcommand{\bea}{\begin{eqnarray}}
\newcommand{\eea}{\end{eqnarray}}

\title{Theory of Photon Subtraction for Two-Mode Entangled Light Beams}

\author[${1}$]{Oscar Rosas-Ortiz}
\author[${2}$]{Kevin Zelaya}

\affil[${1}$]{\footnotesize Physics Department, Cinvestav, AP 14-740, 07000
M\'exico City, Mexico}

\affil[${2}$]{\footnotesize The Czech Academy of Science, Nuclear Physics Institute, 250 68 $\check{R}$e$\check{z}$, Czech Republic}

\date{}
\begin{document}

\maketitle

\begin{abstract}
{\footnotesize Photon subtraction is useful to produce nonclassical states of light addressed to applications in photonic quantum technologies. After a very accelerated development, this technique makes possible obtaining either single photons or optical cats on demand. However, it lacks theoretical formulation enabling precise predictions for the produced fields. Based on the representation generated by the two-mode $SU(2)$ coherent states, we introduce a model of entangled light beams leading to the subtraction of photons in one of the modes, conditioned to the detection of any photon in the other mode. We show that photon subtraction does not produce nonclassical fields from classical fields. It is also derived a compact expression for the output field from which the calculation of conditional probabilities is straightforward for any input state. Examples include the analysis of squeezed-vacuum and odd-squeezed states. We also show that injecting optical cats into a beam splitter gives rise to entangled states in the Bell representation.}
\end{abstract}


\section{Introduction}

Nonclassical states of light are very useful to develop photonic quantum technologies \cite{Obr09}. Considerable attention has been devoted to photon-number states $\vert n \rangle$ since they permit the realization of quantum communication in several forms \cite{Gis07}: 1 photon to send messages, 2 photons to prepare a given quantum state at a distance, 3 photons to teleport quantum states, and so on. Quite remarkably, scalable quantum computing is possible by using single photon sources \cite{Kni01}. In the same context, as the squeezed states of light have less noise in one of their quadratures than the quantum noise limit dictates \cite{Wal83}, they are useful to improve the precision of interferometric measurements dealing with very low intensity light signals \cite{Ros19,Bar20}. The prototypical example is the squeezed-vacuum state, which consists entirely of even-photon number states \cite{Lvo15}. The counterpart of squeezed-vacuum, called odd-squeezed state, includes odd-photon number states only \cite{Zel21}. Additionally, the even and odd coherent states \cite{Dod74} are constructed as opposite phase superpositions of the fully coherent states introduced by Glauber \cite{Gla07}. This issue is of great interest because the Glauber states, formed by superpositions of photon-number states, tolerate a description in terms of the Maxwell theory \cite{Ros19}, so they may describe the states of macroscopic systems. The creation of cat states, a theoretical description introduced by Schr\"odinger to show the way in which quantum mechanics contradicts our everyday experience for systems as great as a `cat' \cite{Sch35}, is therefore available in the laboratory using even and odd coherent states \cite{Aga13,Bac19}, which are therefore called {\em optical cat states} \cite{Zha21} (although the term {\em optical kitten} is also found \cite{Our06}). 

Optical cat states can be experimentally prepared by photon subtraction from a squeezed vacuum state \cite{Our06,Daj97,Nee06,Wak07,Ger10,Nam10,Asa17} and find immediate applications in diverse photonic quantum technologies, including quantum information \cite{Ral03} and quantum teleportation \cite{van01}. Their nonclassical properties have been analyzed in terms of the sub-Poissonian statistics and the negativity of the Wigner function \cite{Kim05,Bis07}, where it has been shown that single-photon subtracted squeezed states decay to vacuum \cite{Bis07}. The optical cat states are also instrumental in the study of entanglement \cite{Cla02,She13,Zha21b,Yan22}.

Photon subtraction offers a practical way of getting single photons on demand from weakly squeezed vacuum \cite{Our06,Ger10,Nam10,Nee06,Wen04,Oli05}. The first antecedents can be traced back to the study of two-mode electromagnetic fields expressed as superpositions of SU(2) coherent states \cite{San89}, including the development of quantum lithography \cite{Bot00,Kok01}, where the occurrence of NOON states \cite{Lee02} is quite natural \cite{Kok02}. The primordial NOON state, with $N=2$ is the result of the Hong-Ou-Mandel effect \cite{Hon85,Hon87}. The production of NOON states for higher values of N was proposed in \cite{Kok02} via conditioned photo detection. Experimental success for $N \leq  5$ has been reported by using different techniques in e.g. \cite{Pry03,Wal04,Afe10,Isr12}.

A very practical version of the photon subtraction technique considers beam splitters to generate two-mode entangled light beams. The process changes the quantum statistics of the input fields so that the output modes are correlated in nonclassical form. Therefore, the field in one of the output modes is conditioned to the result of measuring a given number of photons in the other mode. Formally, lossless beam splitters are associated with the symmetries of the $SU(2)$ Lie group \cite{Cam89}. Indeed, the $SU(2)$ coherent states \cite{Per86} may be represented in two-mode form \cite{Wod86,Buz89,San89,Mor21}, so they coincide with the output of a 50/50 beamsplitter that is injected with $n$-photons in one input channel and $m$-photons in the other input channel \cite{Kim02}. It has been shown that, when the two input modes have the same number of photons, the output state includes a superposition of even photon-number states only \cite{Kim02}, so the possibility of having odd numbers of photons is zero \cite{Lai91}. Besides, the entanglement properties of the output fields are strengthened if the input modes contain different number of photons \cite{Kim02}. The latter is usually  considered by injecting a superposition of photon-number states into one of the input channels and vacuum in the other one \cite{Bis07}.

In this work we provide a theoretical approach for photon subtraction in two-mode entangled light beams. The model considers a lossless symmetric beam splitter to generate nonclassical correlations between the output modes. The main idea is to represent the output fields as linear superpositions of  $SU(2)$ coherent states that exhibit nonclassical properties. We therefore find the conditions for non-separability in the output fields, which are intimately connected with the quantum properties of the input state. In particular, we show that photon subtraction does not produce nonclassical fields from classical fields. We derive a compact expression for the output field from which the calculation of conditional probabilities is straightforward. Our purpose is aimed at making up for the lack of theoretical formulations of the photon subtraction technique, which has been developed mainly in experimental form.

The structure of the paper is as follows. In Sec.~\ref{chap2} we revisit the two-mode representation based on the $SU(2)$ coherent states and discuss the generalities of the two-mode states that can be constructed in terms of such basis. In Sec.~\ref{chap3} we introduce the structure of photon-subtracted states associated with any input field consisting of photon-number state superpositions. We discuss about the conditions for photon subtraction and derive a compact expression for the two-mode entangled fields produced by a beam splitter. In Sect.~\ref{chap4} we show the applicability of our model generating photon-subtracted versions of the squeezed-vacuum and the odd-photon squeezed states. We also study the result of injecting optical cat states into the beam splitter, which leads to entangled states in the form of the Bell-basis elements. A short appendix includes calculations that are useful to reproduce the results of the main text.


\section{Two-mode entangled states}
\label{chap2}

The states of two-mode light beams are elements of the separable Hilbert space ${\cal H} = \operatorname{span} \{ \vert n,m \rangle, n,m =0,1, \ldots \}$, where  the bipartite states $\vert n \rangle_a \otimes \vert m \rangle_b =\vert n, m \rangle$ are orthonormal, with $\vert n \rangle_a$ and $\vert m \rangle_b$ forming the orthonormal bases of the space of photon-number states for modes $a$ and $b$, respectively. A given element $\vert \varphi \rangle \in \mathcal{H}$ is separable if there exists a pair of vectors $\vert \phi \rangle_a$ and $\vert \theta \rangle_b$ such that $\vert \varphi \rangle = \vert \phi \rangle_a \otimes \vert \theta \rangle_b$. Otherwise, $\vert \varphi \rangle$ is said to be non-separable or entangled.

Next, we provide a formulation to construct a new basis for $\mathcal{H}$ that is expressed as non-separable combinations of $\vert n,m \rangle$. The main interest is to facilitate the construction and analysis of two-mode entangled states in $\mathcal{H}$, such that they describe the outputs of a beam splitter and the basic ingredient for photon subtraction as well.

\subsection{Bipartite space of states}

The $SU(2)$ coherent states \cite{Per86} can be constructed in the two mode Hilbert space ${\cal H}$, the representation of which is obtained by applying the unitary operator
\be
\hat U (\xi) = \exp \left[ \xi \hat a^{\dagger} \hat b - \xi^* \hat a \hat b^{\dagger}  \right], \quad \xi = \vert \xi \vert e^{i \varphi}, \quad \varphi \in [-\pi, \pi),
\label{u}
\ee
to the bipartite state $\vert n,0 \rangle$, where $\hat a$, $\hat a^{\dagger}$, and $\hat b$, $\hat b^{\dagger}$, are the boson ladder operators for modes $a$ and $b$ respectively \cite{Wod86,Buz89,San89}.

The operator $\hat U(\xi)$ may be used to represent a lossless symmetric beam splitter, with amplitudes of transmission and reflection $t = \cos \vert \xi \vert$ and $r = \sin \vert \xi \vert$ \cite{Cam89}. The parameter $\varphi$ denotes a phase difference between the transmitted and reflected beams. Then, $\hat U(\xi) \vert n, m \rangle$ is the result of injecting $n$ photons into  channel $a$ and $m$ photons into channel $b$ of the beam splitter. The output state $\hat U(\xi) \vert n, m \rangle$ exhibits entanglement properties that can be strengthened if $n \neq m$ \cite{Kim02}, which is precisely the case for the two-mode $SU(2)$ coherent states $\hat U (\xi)  \vert n, 0 \rangle$.

In this paper we consider an idealized $50/50$ beam splitter $\hat B$, represented by $\hat U(\xi)$ with $\xi = i \frac{\pi}{4}$. That is
\be
\hat B = \exp \left[ i \frac{\pi}{4} \left( a^{\dagger} b + a b^{\dagger} \right) \right].
\label{B}
\ee
The corresponding two-mode $SU(2)$ coherent states acquire the form \cite{Kim02}
\be
\vert n, 0 \rangle_B = \hat B \vert n, 0 \rangle =  \sum_{k=0}^n c_{k, n-k}
 \vert k, n-k \rangle, \qquad c_{k, n-k} = \frac{1}{\sqrt{2^n}} \left( 
\begin{array}{c}
n\\k
\end{array}
\right)^{1/2} e^{i \frac{\pi}{2} k}.
\label{su2}
\ee
The set $\vert n, 0 \rangle_B$ is orthonormal, so it defines a concrete representation in the two-mode Hilbert space ${\cal H}$; the first elements  are provided in Eq.~(\ref{basis}) of Appendix~\ref{ApA}. The change of basis between the sets $\vert n,0 \rangle_B$ and $\vert s,m \rangle$ is ruled by the unitary operator (\ref{B}). The advantage of working in the representation $\vert n,0 \rangle_B$ is that the basis elements are non-separable when they are expressed in representation $\vert s,m \rangle$. The latter means that $\vert n, 0 \rangle_B$ encodes nonclassical correlations between mode $a$ and mode $b$, with exception of $\vert 0, 0 \rangle_B$. 

As the result of injecting $n$-photons in channel $a$ of the beam splitter $\hat B$, and 0-photons in channel $b$, the basis $\vert n, 0 \rangle_B$ shows very useful properties. In particular, finding $m$ photons in mode $a$ implies $r=n-m$ photons in mode $b$. Thus, for any state (\ref{su2}) we have the conditional probability
\be
\mathcal{P}_{m,r}= \frac{1}{2^n} \left( 
\begin{array}{c}
n\\k
\end{array}
\right)
 = \frac{\Gamma(r+\frac12) \Gamma(m+\frac12)}{\Gamma(r+1) \Gamma(m+1)} \left[ \frac{1}{ 2^{r+m} B(r+ \frac12, m +\frac12)} \right], \quad n=m+r,
 \label{proba}
\ee
where $B(x,y)= \frac{\Gamma(x) \Gamma(y)}{\Gamma(x+y)}$ is the Euler beta function \cite{Olv10}, which cannot be factorized as  $B(x,y) = f(x) g(y)$ for any functions $f$ and $g$. Therefore, $\mathcal{P}_{m,r} \neq \mathcal{P}_m \mathcal{P}_r$, with $\mathcal{P}_m$ and $\mathcal{P}_r$ two independent probability distributions, one for each output port of the beam splitter. This property is concomitant to the impossibility of writing the $SU(2)$ coherent states (\ref{su2}) as the product of two independent mode states. That is, if $n\neq m$, then $\mathcal{P}_{m,r} \neq \mathcal{P}_m \mathcal{P}_r$ implies $\vert n, 0 \rangle_B \neq \vert \phi \rangle_a \otimes \vert \theta \rangle_b$ \cite{Zel17}. The latter is a direct consequence of the nonclassical correlations between modes $a$ and $b$ that are encoded in states $\vert n, 0 \rangle_B$. 

\subsection{Representations in the bipartite state space}

 For any regular vector $\vert \Psi \rangle \in {\mathcal H}$ one has
\be
\vert \Psi \rangle_B = \frac{1}{\lambda } \sum_{n=0}^{\infty} \alpha_n \vert n,0 \rangle_B = \frac{1}{\lambda } \sum_{n=0}^{\infty} \sum_{k=0}^n \alpha_n c_{k, n-k} \vert k, n-k \rangle, \quad \alpha_n \in \mathbb C,
\label{state}
\ee
where $\lambda \in \mathbb C$ stands for normalization. Noticeably, we may also write 
\be
\vert \Psi \rangle_B = \hat B \vert \psi(\alpha),  0 \rangle,
\label{stateb}
\ee
with
\be
\vert \psi (\alpha) \rangle_a = \frac{1}{\lambda (\alpha) } \sum_{n=0}^{\infty} \alpha_n \vert n \rangle_a
\label{input1}
\ee
a normalized linear combination of number states in the $a$ mode. That is, we can construct regular two-mode states $\vert \Psi \rangle_B$ by injecting regular superpositions $\vert \psi(\alpha) \rangle_a$ into the $a$-port of the beam splitter $\hat B$. 

For the sake of simplicity, in Eqs.~(\ref{stateb}) and (\ref{input1}) we have introduced the shortcut notation $\vert \psi(\alpha) \rangle := \vert \psi(\alpha_1, \alpha_2, \ldots ) \rangle$. Consistently, $\lambda (\alpha) := \lambda (\alpha_1, \alpha_2, \ldots )$. This notation is adopted for similar expressions throughout the manuscript.

The quantum properties of state $\vert \Psi \rangle_B$ depend on the combined amplitude probabilities $\alpha_n c_{k, n-k}$, which may lead to either separable ({\em classical\/}) or non-separable ({\em nonclassical\/}) states in $\mathcal{H}$, see the discussion of Eq.~(\ref{proba}) above. That is, although the two-mode basis elements $\vert n, 0 \rangle_B$ encode entanglement between modes $a$ and $b$ (with exception of $\vert 0,0 \rangle_B$), their linear superpositions $\vert \Psi \rangle_B$ may lack such entanglement. This property is very common in quantum physics \cite{Ros19}, with the coherent states $\vert z \rangle$ of Glauber \cite{Gla07} as prototypical example. Indeed, decomposing $\vert z \rangle$ into a pair of superpositions, one consisting entirely of even-photon number states and the other including odd-photon number states only, one arrives at the even and odd coherent states of Dodonov, Malkin and Man'ko \cite{Dod74}, which are nonclassical \cite{Dod03}. Hence, the fully coherent (classical) state $\vert z \rangle$ is a superposition of two nonclassical states while the even and odd coherent states are superpositions of two classical states!

We are mainly interested in nonclassical two-mode states $\vert \Psi \rangle_B$, since the quantum correlations between modes $a$ and $b$ are fundamental to construct photon-subtracted states successfully. 

\section{Conditions for photon subtraction}
\label{chap3}

The straightforward calculation shows that the two-mode state (\ref{state}) can be rewritten as follows
\be
\vert \Psi  \rangle_B= \frac{1}{\lambda} \sum_{n=0}^{\infty} 
\frac{e^{i\frac{\pi}{2}n}}{\sqrt{n!}} \lambda_n (\beta) \vert n \rangle_a 
\otimes \vert \psi_n(\beta) \rangle_b,
\label{state3}
\ee
where $\vert \psi_n(\beta) \rangle$ and $\lambda_n(\beta)$ are written in the shortcut notation introduced above, with $\beta_n = \frac{\alpha_n}{\sqrt{2^n}}$, and
\be
\vert \psi_n (\alpha) \rangle_a= \frac{1}{\lambda_n (\alpha)} \sum_{k=0}^{\infty} \sqrt{ \frac{(k+n)!}{k!}} \alpha_{k+n} \vert k \rangle_a,
\label{sus1}
\ee
see details in Appendix~\ref{ApA}.

The vector $\vert \psi_n (\alpha)\rangle_a$ results of applying $n$-times the annihilator operator $\hat a$ on the input state (\ref{input1}), so it represents the subtraction of $n$ photons from $\vert \psi (\alpha) \rangle_a$. Hereafter this vector will be referred to as the $n$-subtracted state of $\vert \psi (\alpha) \rangle_a$. 

As $\vert \psi_n (\beta) \rangle_b$ in (\ref{state3}) is the $b$ mode version of (\ref{sus1}), evaluated with the reduced probability amplitudes $\beta_n =\alpha_n /\sqrt{2^n}$, and the beam splitter $\hat B$ is lossless and symmetric, the output (\ref{state3}) can be also written in the form
\[
\vert \Psi  \rangle_B= \frac{1}{\lambda} \sum_{k =0}^{\infty} 
\frac{e^{-i\frac{\pi}{2}k}}{\sqrt{k!}} \lambda_k(\beta) \vert \psi_k (\beta) \rangle_a \otimes \vert k \rangle_b.
\]
Thus, detecting a given number of photons in any of the two output ports, the beam in the other port is represented by a subtracted photon state.

In the sequel we concentrate in representation (\ref{state3}). The output ports $a$ and $b$ of $\hat B$ will be referred to as idler and signal. Consistently, the modes $a$ and $b$ in  $\vert \Psi \rangle_B$ will be respectively called idler and signal. Besides, to generate subtracted photon states in the signal channel, we assume that idealized photodetectors with unit efficiency are used to collect photons at the idler channel.

Depending on $\alpha_n$, the two-mode state (\ref{state3}) will exhibit nonclassical correlations between modes $a$ and $b$. The photon subtraction operates whenever $\vert \Psi \rangle_B$ is not factorized as the product of two independent states, one for mode $a$ and one for mode $b$. Thus, entanglement between idler and signal output channels is necessary to link the occurrence of the $n$-subtracted signal state $\vert \psi_n (\beta) \rangle_b$ with the detection of exactly $n$-idler photons. 

According to the conjecture that entangled output states from a beam splitter require nonclassicality in the input port \cite{Kim02}, the first clue to produce photon subtraction is to consider nonclassical states $\vert \psi (\alpha) \rangle_a$. Another trail is reached by noticing that a very special class of probability amplitudes fulfilling 
\be
\alpha_{k+n} = \frac{1}{\sqrt{(k+n)!}} \delta_k \gamma_n 
\label{separa}
\ee 
lead to separable summations in (\ref{state3}), and then to separable versions of the output state.
Indeed, introducing (\ref{separa}) into the pair of equations (\ref{sus1}) and (\ref{state3}) yields
\be
\vert \Psi \rangle_B = \tfrac{\lambda(\widetilde \delta) \lambda(\widetilde \gamma)}{\lambda} \vert \psi(\widetilde \gamma) \rangle_a \otimes \vert \psi (\widetilde \delta) \rangle_b, \quad \widetilde \gamma_n = \frac{e^{i \frac{\pi}{2} n} \gamma_n}{\sqrt{2^n}}, \quad \widetilde \delta_k = \frac{\delta_k} {\sqrt{2^k}}.
\label{separa2}
\ee
The output field (\ref{separa2}) is separable, so it does not encode nonclassical correlations between modes $a$ and $b$. Besides, following \cite{Kim02}, the input field constructed with the probability amplitudes (\ref{separa}) is classical by necessity. Thus, we have shown that photon subtraction of classical fields does not produce nonclassical fields.

Probability amplitudes fulfilling (\ref{separa}) give rise to independent probabilities for detecting photons in either of the output channels. They are intimately connected with the separability of bipartite states \cite{Zel17}, so their identification is complementary to the conjecture introduced in \cite{Kim02}. 

\subsection{Classical correlations}

The simplest form to obtain factorized states (\ref{state3}) involves scalable probability amplitudes like $\alpha_n = \frac{z^n}{\sqrt{n!}}$, since (\ref{separa}) is fulfilled with $\gamma_n= z^n$ and $\delta_k =z^k$. In this case the input (\ref{input1}) is a Glauber state \cite{Gla07}:
\be
\vert z \rangle =  \frac{1}{\lambda_G} \sum_{n=0}^{\infty}   \alpha^G_n \vert n \rangle, \quad \lambda_G(z) = e^{\frac{\vert z \vert^2}{2}}, \quad \alpha^G_n= \frac{z^n}{\sqrt{ n!}}.
\label{Glau}
\ee
Then, Eq.~(\ref{state3}) is reduced to the factorized form (\ref{separa2}). Explicitly
\be
\vert Z \rangle_B= \left\vert \frac{iz}{\sqrt 2} \right\rangle_a \otimes \left\vert \frac{z}{\sqrt 2} \right\rangle_b = \left\vert \frac{iz}{\sqrt 2},  \frac{z}{\sqrt 2} \right\rangle.
\label{Glau2}
\ee
Notably, the reduced probability amplitude $\beta_n$ coincides in form with $\alpha_n$, but changing the complex parameter $z$ by $\frac{1}{\sqrt 2} z$. Recalling that the expected value for the number of photons $\hat n$ in a Glauber state is $\langle \hat n \rangle= \vert z \vert^2$, we see that the factors of $\vert Z  \rangle_B$ are Glauber states with $\frac12 \langle \hat n \rangle$. The latter is consistent with the behavior of Gaussian laser beams that are injected into a beam splitter. For actual beam splitters ($\approx 50/50$), the output is a pair of Gaussian beams with intensity that is approximately one half the input intensity. Moreover, detecting a photon in either of the output channels does not affect the nature of the field in the other channel. The above properties are explained by the fact that the output system does not include quantum correlations between their components. The result may be considered classical in two forms: On the one hand, the probability $\mathcal{P}_{n,m}$ of detecting $n$-idler photons and $m$-signal photons can be expressed as the product of two independent Poisson distributions with mean value $\frac{\vert z \vert^2}{2}$, which is a fingerprint of classicalness \cite{Zel17}. On the other hand, each of the output modes is a fully coherent state,  so they are classical in the sense that tolerate a description in terms of the Maxwell theory \cite{Ros19}. Thus, the output state (\ref{Glau2}) verifies that the photon subtraction of classical fields does not produce nonclassical fields.

\subsection{Non-classical correlations}

Assuming that (\ref{state3}) is not separable, the number of idler photons will determine the $n$-subtracted signal state $\vert \psi_n (\beta) \rangle_b$. To be concrete, the expression
\be
\mathcal{P}_n (\alpha) =  \left\vert \frac{\lambda_n (\beta)}{\lambda(\alpha)} \right\vert^2 \frac{1}{n!}
\label{result}
\ee
provides the probability of finding $n$ photons in the idler channel and state $\vert \psi_n (\beta) \rangle_b$ in the signal channel. The behavior of $\mathcal{P}_n (\alpha)$ depends on the concrete analytical expression of $\lambda_n(\beta)$, so the success of subtracting photons from the signal beam is markedly determined by the properties of the input state through the amplitude probabilities $\alpha_n$.

To facilitate applications, the following expressions provide the form acquired by (\ref{state3}) for input states $\vert \psi_E(\alpha) \rangle$ and $\vert \psi_O (\alpha) \rangle$, consisting entirely of even-photons and odd-photons in the superposition (\ref{input1}). Thus
\be
\begin{array}{l}
\vert \Psi_E \rangle_B =\displaystyle\frac{1}{\lambda_E} \sum_{n=0}^{\infty}  
\left[ \frac{e^{i\pi n}}{\sqrt{(2n)!}} \lambda^E_{2n} (\beta) \vert 2n \rangle_a \otimes \vert \psi^E_{2n} (\beta) \rangle_b 
\right. \\[3ex]
\hskip5cm
\left.
+ \displaystyle \frac{i e^{i\pi n}}{\sqrt{(2n+1)!}} \lambda^E_{2n+1} (\beta) \vert 2n+1 \rangle_a \otimes \vert \psi^E_{2n+1} (\beta) \rangle_b 
\right],
\end{array}
\label{parte5}
\ee
and
\be
\begin{array}{l}
\vert \Psi_O \rangle_B = 
\displaystyle\frac{1}{\lambda_O} \sum_{n=0}^{\infty}
\left[
\frac{e^{i \pi n}}{\sqrt{(2n)!}} \lambda_{2n}^O( \beta)
\vert 2n \rangle_a \otimes \vert \psi_{2n}^O( \beta)  \rangle_b
\right. \\[3ex]
\hskip5cm
\left.
+  \displaystyle \frac{i e^{i \pi n}}{\sqrt{(2n+1)!}} \lambda_{2n+1}^O( \beta)
\vert 2n +1 \rangle_a \otimes \vert \psi_{2n+1}^O( \beta)  \rangle_b
\right],
\end{array}
\label{parte6}
\ee
represent the result of injecting {\em even} states $\vert \psi_E(\alpha) \rangle$ and {\em odd} states $\vert \psi_O(\alpha) \rangle$ into the beam splitter, respectively. Concrete expressions for the $n$-subtracted states of $\vert \psi_E (\alpha)\rangle$ and $\vert \psi_O (\alpha) \rangle$ are given in Eqs.~(\ref{sus2even})-(\ref{sus2odd}) and (\ref{sus3even})-(\ref{sus3odd}) of Appendix~\ref{ApA}.


\section{Applications}
\label{chap4}

Equation~(\ref{state3}) comprises information that is necessary to describe the photon subtraction  (\ref{input1}) of an input state $\vert \psi (\alpha) \rangle$. The related output $\vert \Psi \rangle_B$  is a linear superposition of entangled bipartite states whose probability amplitudes yield the probability (\ref{result}). Each element in the superposition links a photon number state $\vert n \rangle_a$ of mode $a$ with the $n$-subtracted state (\ref{sus1}) of $\vert \psi (\alpha) \rangle$ in mode $b$. Additionally, Eqs.~(\ref{parte5}) and (\ref{parte6}) provide the above formulae assuming that the input state $\vert \psi (\alpha) \rangle$ is composited only of either even-number states or odd-number states, respectively. All these results are useful to further analyze properties like the mean value of dynamical variables, photon-statistics, and the negativity of the corresponding Wigner function.  

Next we apply our method to study photon subtraction in three different cases that are of interest.

\subsection{Squeezed-vacuum state}
\label{vacio}

We consider the squeezed-vacuum state 
\be
\vert \xi^{\operatorname{vac}} \rangle=  (1-\vert \xi \vert^2)^{1/4} \sum_{n=0}^{\infty} 
\frac{\sqrt{(2n)!}}{n!} \left( - \frac{\xi}{2} \right)^n \vert 2 n \rangle, \quad \vert \xi \vert < 1,
\label{vacuum}
\ee
The beam splitter output $\vert \xi^{\operatorname{vac}},0 \rangle_B$ is easily calculated from (\ref{parte5}) and the $n$-subtracted states
\be
\vert \xi^{\operatorname{vac}}_{2n} \rangle =
\frac{1}{\lambda_{2n}^{\operatorname{vac}} (\xi)} 
\sum_{k=0}^{\infty} \frac{1}{\sqrt{(2k)!}}  \frac{(2k+2n)!}{(k+n)!}
\left( - \frac{\xi}{2} \right)^{k+n} \vert 2 k \rangle,
\label{ppar3}
\ee

\be
\vert \xi^{\operatorname{vac}}_{2n+1} \rangle = 
\frac{1}{\lambda_{2n+1}^{\operatorname{vac}} (\xi)}
\sum_{k=0}^{\infty} \frac{1}{\sqrt{(2k+1)!}}  \frac{(2k+2n+2)!}{(k+n+1)!}
\left( - \frac{\xi}{2} \right)^{k+n+1} \vert 2 k +1\rangle.
\label{podd3}
\ee
The normalizations are defined by the formulae
\bea
\vert \lambda_{2n}^{\operatorname{vac}}  (\xi)  \vert^2 =   \left( \frac{\vert \xi \vert}{2} \right)^{2n}
\left[ \frac{\Gamma(2 n+1)}{\Gamma(n+1)} \right]^2
{}_2F_1(n +\tfrac12, n +\tfrac12, \tfrac12;  \vert \xi \vert^2 ),
\\[3ex]
\vert \lambda_{2n+1}^{\operatorname{vac}} (\xi) \vert^2 = \left( \frac{\vert \xi \vert}{2} \right)^{2n+1} 
\left[ \frac{\Gamma(2n+3)}{\Gamma(n+2)} \right]^2
{}_2F_1(n+\tfrac32, n +\tfrac32, \tfrac32; \vert \xi \vert^2 ).
\eea
The above expressions are in full agreement with the results already reported in \cite{Bis07}, where special attention is payed to the 1-subtracted state $\vert \xi^{\operatorname{vac}}_1 \rangle$ derived from (\ref{podd3}). 

We would like to emphasize that, according to Eq.~(\ref{parte5}), the two-mode state $\vert \xi^{\operatorname{vac}},0 \rangle_B$ is expressed as a linear combination of $\vert \xi^{\operatorname{vac}}_{2n} \rangle$ and $\vert \xi^{\operatorname{vac}}_{2n+1} \rangle$, calculated in the $\beta$-configuration defined by the reduced probability amplitudes $\beta_k = \alpha_k/\sqrt{2^k}$, with $\alpha_k$ the probability amplitudes of (\ref{ppar3}) and (\ref{podd3}), respectively. The straightforward calculation shows that $\vert \xi^{\operatorname{vac}},0 \rangle_B$ is therefore parameterized by $\frac{\xi}{\sqrt{2}}$, with $\xi$ the complex-number characterizing the squeezed-vacuum state (\ref{vacuum}). Thus, even if no photons are detected at channel $a$, the signal field is different from the input beam (\ref{vacuum}). The latter is clear by recalling that $\xi$ characterizes predictions like the mean value of the number of photons. This phenomenon is explained by the entanglement of the basis elements $\vert n, 0 \rangle_B$, which is preserved by the probability amplitudes defining Eq.~(\ref{vacuum}). That is, also the idler vacuum-state $\vert 0 \rangle_a$ is correlated with the signal field in nonclassical form. As a consequence, the input field is always affected when passing through the beam splitter. 

In optics, it is well known that beam splitters reduce the intensity of light beams \cite{Bac19,Aga13}. Indeed, as the transmission coefficient $\vert t \vert^2$ is always less than one, the field of the transmitted field is reduced by a factor $t $. Hence, ``if observation is made only on the transmitted beam, the beam splitter is just an attenuator of the light beam'' \cite{Aga13}. Consistently, detecting 0-photons at the idler channel, we obtain an attenuated version $\vert \xi^{\operatorname{vac}}_0 \rangle$ of the input field $\vert \xi^{\operatorname{vac}} \rangle$ in the signal mode. 

As we are looking for subtraction of photons in the signal mode, detecting 0-photons in idler mode is the ``no success'' event, which in turn is the most probable.

\begin{figure}[htb]
\centering
\subfloat[][0-photon]{\includegraphics[width=.3\textwidth]{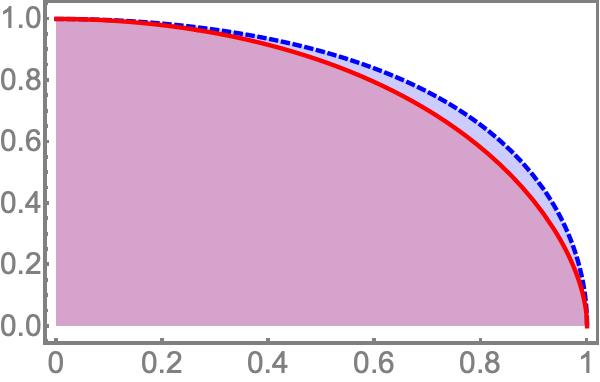}} 
\hskip4ex
\subfloat[][1-photon]{\includegraphics[width=.3\textwidth]{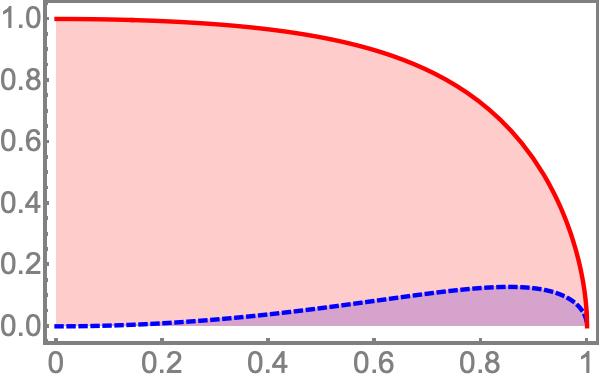}}
\hskip4ex
\subfloat[][2-photon]{\includegraphics[width=.3\textwidth]{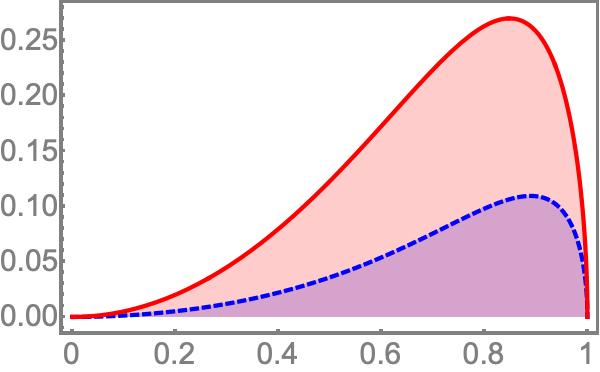}}

\caption{\footnotesize (color online) Probability of photon subtraction for squeezed-vacuum $\vert \xi^{\operatorname{vac}} \rangle$ and odd-photon squeezed $\vert \xi^{\operatorname{odd}} \rangle$ states, blue-dotted and red curves respectively. The horizontal axis corresponds to $\vert \xi \vert$. In both cases the probability of no success is very high, which justifies the technical difficulties to prepare photon-subtracted states in laboratory. Also in both cases the probability of success decreases as the number of subtracted photons increases.  
}
\label{probable}
\end{figure}

Figure~\ref{probable} shows in dotted-blue the probability of finding $n$-subtracted states of $\vert \xi^{\operatorname{vac}} \rangle$ in signal mode. As indicated above, the unsuccessful subtraction of photons ($n=0$) is the highest possible result. In comparison, the probability of success ($n \geq1$) is drastically reduced, which justifies the technical difficulties to prepare photon-subtracted versions of the squeezed-vacuum  in laboratory. 

Interestingly, the above results may be managed to produce single photons on demand. Making $\vert \xi \vert \ll 1$, the input state $\vert \xi^{\operatorname{vac}} \rangle$ can be expanded up to the first photon-number states as follows. Recall that $\vert \xi^{\operatorname{vac}} \rangle$ results from the application of the squeezing operator
\be
\hat S(\xi) = \exp \left[ \frac{\xi}{2} a^{\dagger 2} - \frac{\xi^*}{2} a^2\right]
\label{S}
\ee
on the vacuum state $\vert 0 \rangle$. The series expansion yields
\[
\hat S(\xi) = \mathbb{I} + \frac{\xi}{2} a^{\dagger 2} - \frac{\xi^*}{2} a^2 + \frac12 \left[ \frac{\xi}{2} a^{\dagger 2} - \frac{\xi^*}{2} a^2 \right]^2 + \cdots
\]
Then, up to first order in $\vert \xi \vert$, one gets $\vert \xi^{\operatorname{vac}} \rangle = \hat S(\xi) \vert 0 \rangle \approx \vert 0 \rangle + \frac{\xi}{\sqrt{2}} \vert 2 \rangle$. The action of the beam splitter on the latter approximate state gives
\[
\vert \xi^{\operatorname{vac}},0 \rangle_B  \approx \vert 0,0 \rangle_B + \frac{\xi}{\sqrt{2}} \vert 2,0 \rangle_B =\vert 0,0 \rangle + \frac{\xi}{\sqrt{2}} \left( \frac12 \vert 0,2 \rangle + \frac{i}{\sqrt 2} \vert 1,1 \rangle - \frac12 \vert 2, 0 \rangle
\right).
\]
So that
\be
\vert \xi^{\operatorname{vac}},0 \rangle_B  \approx 
\vert 0 \rangle_a \otimes \left( \vert 0 \rangle_b + \frac{\xi}{2\sqrt 2} \vert 2 \rangle_b \right) + i\frac{\xi}{2} \vert 1 \rangle_a \otimes \vert 1 \rangle_b - \frac{\xi}{2 \sqrt 2} \vert 2 \rangle_a \otimes \vert 0 \rangle_b
\label{series1}
\ee
predicts the production of one single photon in channel $b$ by detecting one photon in channel $a$. 

\begin{figure}[htb]
\centering
\subfloat[][$p=0$]{\includegraphics[width=.3\textwidth]{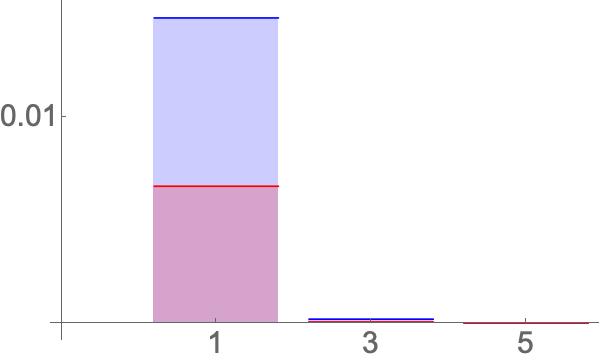}} 
\hskip4ex
\subfloat[][$p=1$]{\includegraphics[width=.3\textwidth]{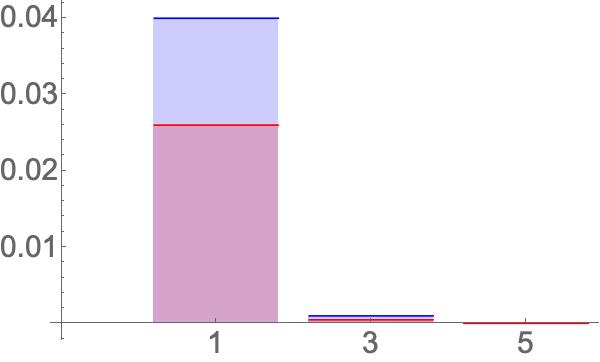}} 
\hskip4ex
\subfloat[][$p=2$]{\includegraphics[width=.3\textwidth]{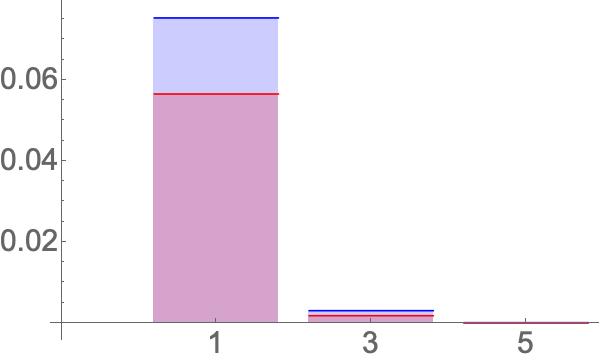}}

\caption{\footnotesize (color online) Photon distribution for the $(2p+1)$-subtracted squeezed-vacuum $\vert \xi^{\operatorname{vac}}_{2p+1} \rangle$ (blue columns) and the $2p$-subtracted odd-squeezed state $\vert \xi^{\operatorname{odd}}_{2p} \rangle$ (red columns). Both versions of subtracted states are composited by odd-photon number states only, see Table~\ref{table1}. In all cases $\vert \xi \vert =0.1$, and the value of $p$ is indicated in the caption.
}
\label{histo1}
\end{figure}

\begin{figure}[htb]
\centering
\subfloat[][$p=0$]{\includegraphics[width=.3\textwidth]{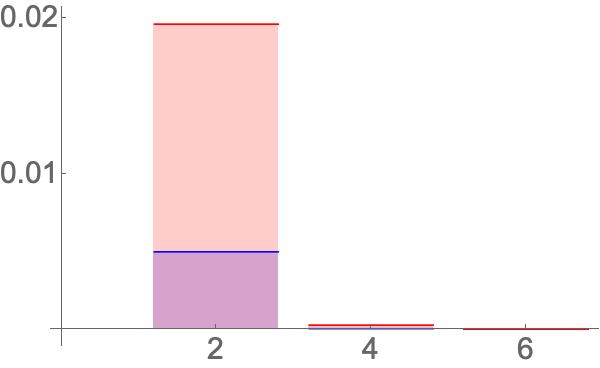}} 
\hskip4ex
\subfloat[][$p=1$]{\includegraphics[width=.3\textwidth]{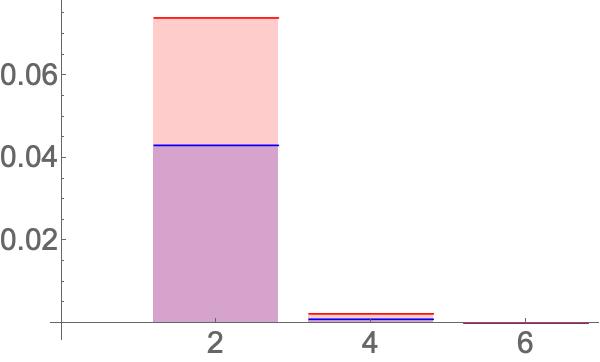}} 
\hskip4ex
\subfloat[][$p=2$]{\includegraphics[width=.3\textwidth]{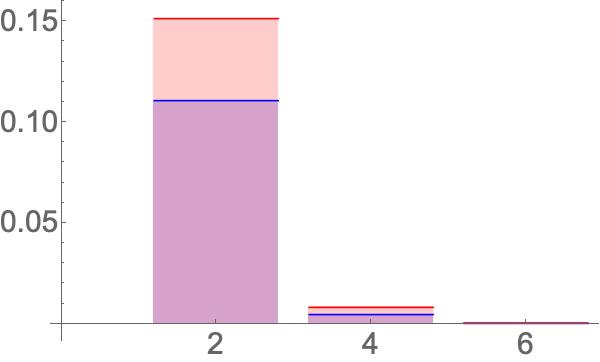}}

\caption{\footnotesize (color online) Photon distribution for the $2p$-subtracted squeezed-vacuum $\vert \xi^{\operatorname{vac}}_{2p} \rangle$ (blue columns) and the $(2p+1)$-subtracted odd-squeezed state $\vert \xi^{\operatorname{odd}}_{2p+1} \rangle$ (red columns). Both versions of subtracted states are composited by even-photon number states only, see Table~\ref{table1}. In all cases $\vert \xi \vert =0.1$, and the value of $p$ is indicated in the caption. Compare with Figure~\ref{histo1}.
}
\label{histo2}
\end{figure}

\begin{figure}[htb]
\centering
\subfloat[][$p=0$]{\includegraphics[width=.3\textwidth]{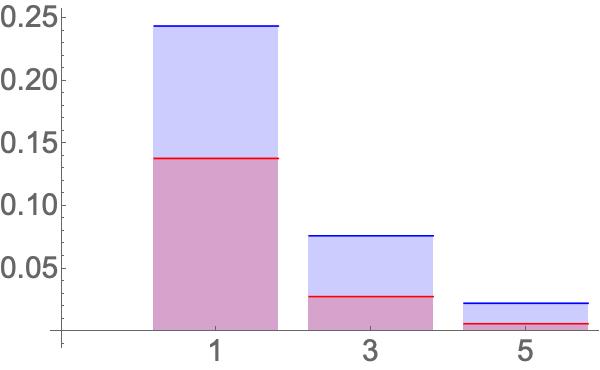}} 
\hskip4ex
\subfloat[][$p=1$]{\includegraphics[width=.3\textwidth]{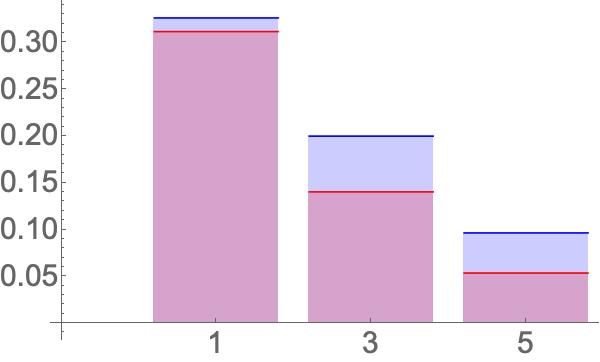}} 
\hskip4ex
\subfloat[][$p=2$]{\includegraphics[width=.3\textwidth]{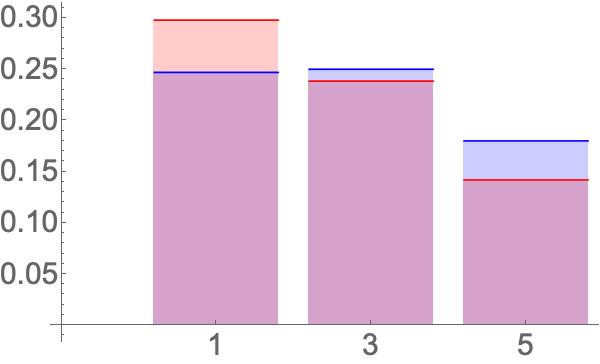}}

\caption{\footnotesize (color online) Same as Figure~\ref{histo1}, with $\vert \xi \vert =0.5$. Comparing with  data in Figure~\ref{histo1}, it is notable the increment of probabilities for photon-number states other than $\vert 1 \rangle$.
}
\label{histo3}
\end{figure}

Figures~\ref{histo1} and \ref{histo2} show respectively the photon distribution for the $n$-subtracted states $\vert \xi^{\operatorname{vac}}_{2n+1} \rangle$ and $\vert \xi^{\operatorname{vac}}_{2n} \rangle$, with $\vert \xi \vert \ll 1$. In both cases it is privileged the production of the photon-number state with the lowest label in the expansion, $\vert 1 \rangle$ and $\vert 2 \rangle$, respectively. Larger values of $\vert \xi \vert$ motivate the increment of probabilities for other photon-number states, see Figure~\ref{histo3}.

\subsection{Odd-photon squeezed states}
\label{oddsec}

In a previous work \cite{Zel21} we have reported a new class of squeezed states that includes only odd-photon number states in their composition. These are called odd-photon squeezed states and are, in many respects, the counterpart of the squeezed-vacuum discussed in the previous section. Indeed, both the squeezed-vacuum and the odd-photon squeezed states satisfy the same second-order difference equation, although they are defined with different initial conditions  \cite{Zel21}. 

To be concrete, the odd-photon squeezed states \cite{Zel21} are given by
\be
\vert \xi^{\operatorname{odd}} \rangle = \frac{1}{\lambda^{\operatorname{odd}}} \sum_{n=0}^{\infty} \frac{n!}{\sqrt{(2n+1)!}} (-2 \xi)^n \vert 2n +1 \rangle, \quad \vert \xi \vert <1,
\label{miodd}
\ee
where
\be
\lambda^{\operatorname{odd}} = \left[ \frac{\vert \xi \vert}{\arcsin \vert \xi \vert}\right]^{-1/2} (1 - \vert \xi \vert^2 )^{-1/4}.
\ee
These vectors may be also written as \cite{Zel21}
\be
\vert \xi^{\operatorname{odd}} \rangle = \frac{1}{\lambda^{\operatorname{odd}}}  {}_1 F_1 (\tfrac 12, \tfrac 32, \tfrac12 \xi \hat a^{\dagger 2}) \hat S(\xi) \vert 1 \rangle,
\label{miodd2}
\ee
with $\hat S(\xi)$ the squeezing operator (\ref{S}). Thus, the odd-photon squeezed states $\vert \xi^{\operatorname{odd}} \rangle$ are the result of applying a confluent hypergeometric-like version of the creation operator $\hat a^{\dagger}$ on the squeezed one-photon state $\hat S(\xi) \vert 1 \rangle$.

Applying the theory of photon subtraction to analyze the odd-photon squeezed state (\ref{miodd}), we obtain the output $\vert \xi^{\operatorname{odd}},0 \rangle_B$ through Eq.~(\ref{parte6}), with
\be
\vert \xi^{\operatorname{odd}}_{2n} \rangle = \frac{1}{\lambda^{\operatorname{odd}}_{2n} (\xi)} \sum_{k=0}^{\infty} \frac{(k+n)!}{\sqrt{(2k +1)!}} (-2 \xi)^{k+n} \vert 2k +1 \rangle
\label{odd1}
\ee
and
\be
\vert \xi^{\operatorname{odd}}_{2n+1} \rangle = \frac{1}{\lambda^{\operatorname{odd}}_{2n+1} (\xi)} \sum_{k=0}^{\infty} \frac{(k+n)!}{\sqrt{(2k )!}} (-2 \xi)^{k+n} \vert 2k \rangle.
\label{odd2}
\ee
The normalizations are defined by the formulae
\bea
\vert \lambda^{\operatorname{odd}}_{2n} (\xi) \vert^2 = 
(2 \vert  \xi \vert)^{2n} \left[ \Gamma(n+1) \right]^2
{}_2F_1(p +1, n +1, \tfrac32;  \vert \xi \vert^2 ),
\\[2ex]
\vert \lambda^{\operatorname{odd}}_{2n+1} (\xi) \vert^2 = 
(2 \vert  \xi \vert)^{2n} \left[ \Gamma(n+1) \right]^2
{}_2F_1(n +1, n +1, \tfrac12;  \vert \xi \vert^2 ).
\eea

Figure~\ref{probable} shows in continuous-red the probabilities to construct the $n$-subtracted states (\ref{odd1})-(\ref{odd2}) by detecting $n$ photons in the idler mode. As in the previous example, no success $(n=0)$ is the most probable event and the probability of success decays for $n >1$. Note however that the probability to get $\vert \xi^{\operatorname{odd}}_1 \rangle$ in the signal channel is relevant for almost any value of $\vert \xi \vert$. The reason is that this state is composited by even-photon number states $\vert 2n \rangle$, just like $\vert \xi^{\operatorname{vac}}_0 \rangle$. Depending on $n$, this similarity is observed for the other $n$-subtracted photon versions of $\vert \xi^{\operatorname{vac}} \rangle$ and $\vert \xi^{\operatorname{odd}} \rangle$, see Table~\ref{table1}. 

\begin{table}
\centering
\begin{tabular}{lclcl}
\hline
\multicolumn{5}{c}{$n$-subtracted photon states}\\
\hline
&&&& \\
input & & $\vert \xi^{\operatorname{vac}} \rangle$ && $\vert \xi^{\operatorname{odd}} \rangle$\\[2ex]
even-like & & $\vert \xi^{\operatorname{vac}}_{2n} \rangle$ && $\vert \xi^{\operatorname{odd}}_{2n+1} \rangle$
\\[2ex]
odd-like & & $\vert \xi^{\operatorname{vac}}_{2n+1} \rangle$ && $\vert \xi^{\operatorname{odd}}_{2n} \rangle$\\[1ex]
\hline
\end{tabular}
\caption{\footnotesize The photon subtraction is available by injecting a light beam $\vert \psi \rangle$ into one of the input channels of a 50/50 beam splitter $\hat B$, and the 0-photon state $\vert 0 \rangle$ into the other. The output $\vert \Psi \rangle_B$ is a two-mode light beam that may exhibit nonclassical correlations between its modes, depending on the quantum properties of the input. Appropriate outputs $\vert \Psi \rangle$ lead to $n$-subtracted states $\vert \psi_n \rangle$ in one channel by measuring $n$ photons in the other channel. The squeezed-vacuum $\vert \xi^{\operatorname{vac}} \rangle$ and odd-photon squeezed $\vert \xi^{\operatorname{odd}} \rangle$ states, entirely composited by even- and odd-photon number states respectively, produce two-mode entangled light beams allowing photon subtraction. In both cases, detection of photons in one of the output channels may produce either even-like or odd-like $n$-subtracted beams at the other output channel.
}
\label{table1}
\end{table}

The odd-squeezed states $\vert \xi^{\operatorname{odd}} \rangle$ are also useful to produce single photons, see Figures~\ref{histo1}, \ref{histo2}, and \ref{histo3}. Indeed, Eq.~(\ref{miodd}) may be rewritten in the form \cite{Zel21}
\be
\vert \xi^{\operatorname{odd}} \rangle = \frac{1}{\lambda^{\operatorname{odd}}} \sum_{n=0}^{\infty} \frac{n!}{(2n+1)!} (-2 \xi a^{\dagger 2})^n \vert 1 \rangle.
\ee
Therefore, using the approximation $\lambda^{\operatorname{odd}}  \vert \xi^{\operatorname{odd}} \rangle \approx \vert 1 \rangle - 2\sqrt{\frac23} \xi \vert 3 \rangle$ we write
\bea
\lambda^{\operatorname{odd}} \vert \xi^{\operatorname{odd}},0 \rangle_B \approx \vert 0 \rangle_a \otimes \left( \tfrac{i}{\sqrt 2} \vert 1 \rangle_b - \tfrac1{\sqrt 3} \xi \vert 3 \rangle_b \right) +  \vert 1 \rangle_a \otimes \left( \tfrac{1}{\sqrt 2} \vert 0 \rangle_b - \xi \vert 2 \rangle_b \right) - 2 \xi \vert 2 \rangle_a \otimes \vert 1 \rangle_b,
\label{series2}
\eea
which predicts the production of one single photon in channel $b$ by detecting two photons in channel $a$, compare with (\ref{series1}). In contraposition with $\vert \xi^{\operatorname{vac}} \rangle$, the above expression shows that making $\vert \xi \vert \ll1$ the series may be further truncated to get one-photon state $\vert 1 \rangle$ in the no success event.

\begin{figure}[htb]
\centering
\subfloat[][odd-like]{\includegraphics[width=.3\textwidth]{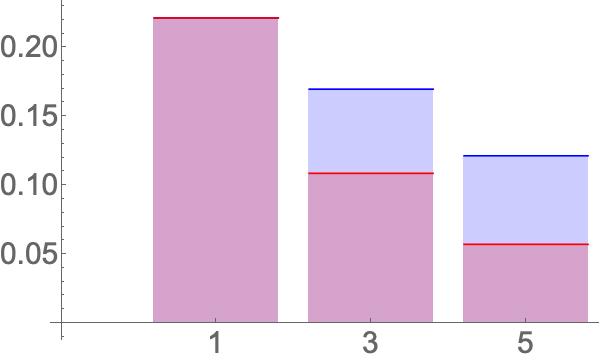}} 
\hskip4ex
\subfloat[][even-like]{\includegraphics[width=.3\textwidth]{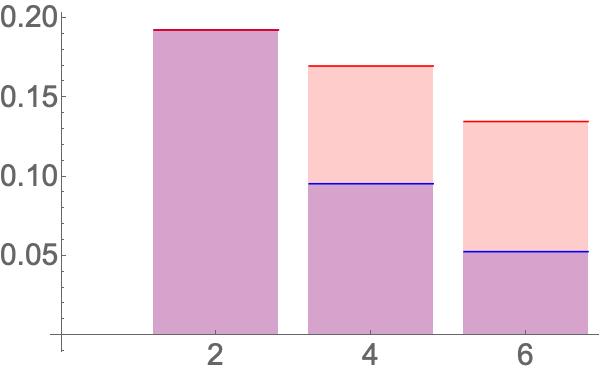}} 

\caption{\footnotesize (color online) Photon distribution for the $n$-subtracted states $\vert \xi^{\operatorname{vac}}_1 \rangle$, $\vert \xi^{\operatorname{odd}}_0 \rangle$, with $\vert \xi \vert = 0.783$ (a) and $\vert \xi^{\operatorname{vac}}_0 \rangle$, $\vert \xi^{\operatorname{odd}}_1 \rangle$, with $\vert \xi \vert = 0.813$ (b). The squeezing parameter has been selected to provide the same probability for states $\vert 1 \rangle$ and $\vert 2 \rangle$, (a) and (b) respectively.
}
\label{histo4}
\end{figure}

On the other hand, controlling the $\xi$-parameter, Figure~\ref{histo4}(a) shows the situation in which the one-photon state $\vert 1 \rangle$ occurs with the same probability for both $\vert \xi^{\operatorname{vac}}_1 \rangle$ and $\vert \xi^{\operatorname{odd}}_0 \rangle$. Note however the prevalence of the former over the latter to produce also states $\vert 3 \rangle$ and $\vert 5 \rangle$. That is, $\vert \xi^{\operatorname{odd}}_0 \rangle$ is more efficient to produce single photons $\vert 1 \rangle$ since it reduces the probabilities to get any other photon-number state. The roles are reversed if one pays attention to state $\vert 2 \rangle$, see Figure~\ref{histo4}(b). In this case, better than $\vert \xi^{\operatorname{odd}}_1 \rangle$, 
the state $\vert \xi^{\operatorname{vac}}_0 \rangle$ reduces the probabilities to get any other state $\vert 2 n \rangle$.

Having two theoretical predictions to produce a given photon-number state, like the ones shown in Figure~\ref{histo4}, may serve to get a better fit between theoretical modeling and experimental data. By manipulating $\xi$, the theoretical photon distributions can be matched to the event of maximal counts, say this corresponds to the occurrence of  $\vert n_0 \rangle$. Then, comparing the theoretical predictions with the counts for other states $\vert n_0 \pm k \rangle$, $n_0 \geq k$, should determine whether $n$-subtracted version of $\vert \xi^{\operatorname{vac}} \rangle$ and $\vert \xi^{\operatorname{odd}} \rangle$ is the best option.

\begin{figure}[htb]
\centering
\subfloat[][$\vert\xi_{0}^{\operatorname{odd}}\rangle$]{\includegraphics[width=0.22\textwidth]{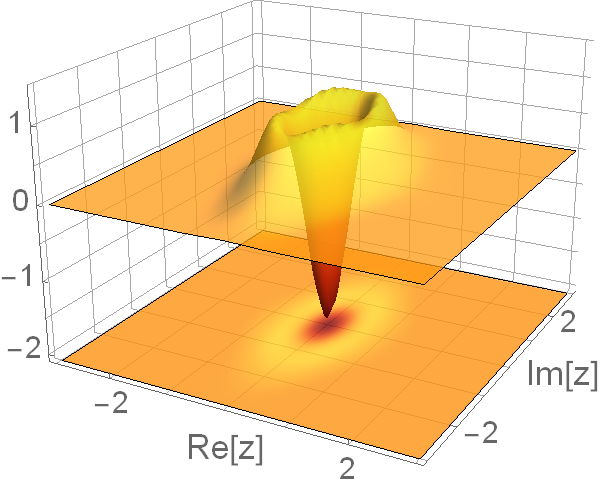}
\label{FWiga}}
\hspace{2mm}
\subfloat[][$\vert\xi_{2}^{\operatorname{odd}}\rangle$]{\includegraphics[width=0.22\textwidth]{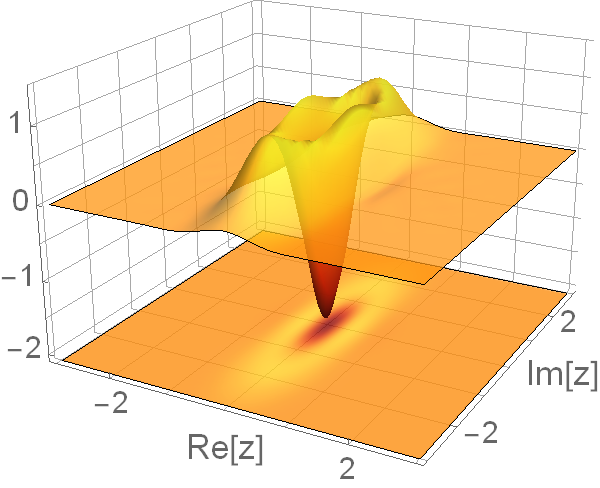}
\label{FWigb}}
\hspace{2mm}
\subfloat[][$\vert\xi_{4}^{\operatorname{odd}}\rangle$]{\includegraphics[width=0.22\textwidth]{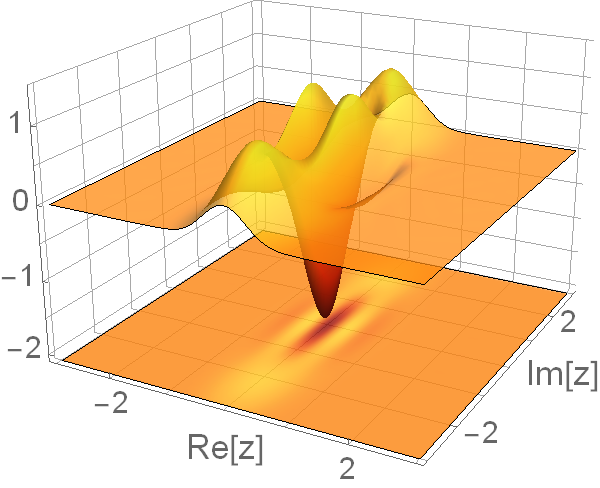}
\label{FWigc}}
\\
\vspace{4mm}
\subfloat[][$\vert\xi_{1}^{\operatorname{odd}}\rangle$]{\includegraphics[width=0.22\textwidth]{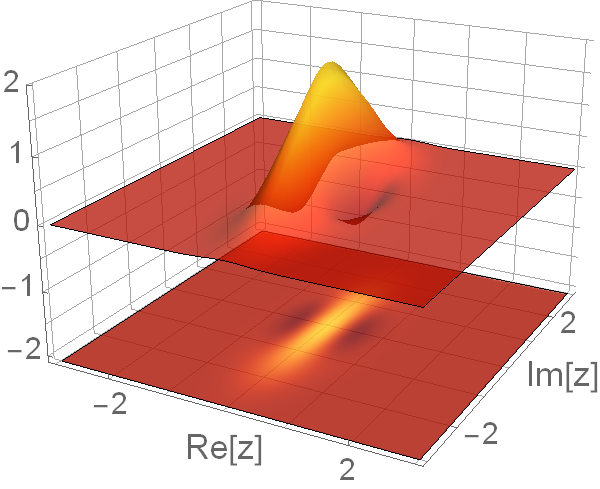}
\label{FWigd}}
\hspace{2mm}
\subfloat[][$\vert\xi_{3}^{\operatorname{odd}}\rangle$]{\includegraphics[width=0.22\textwidth]{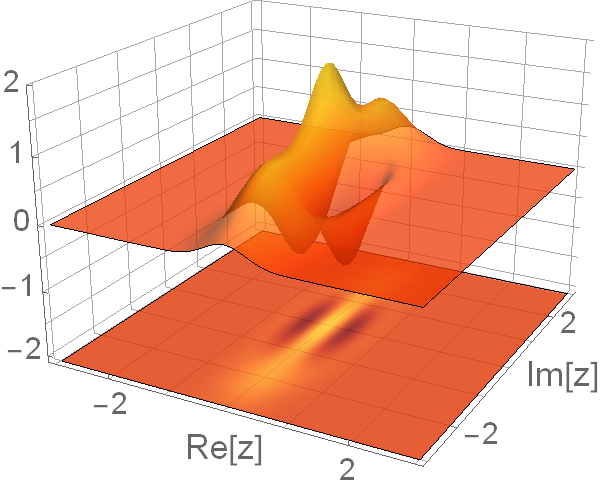}
\label{FWige}}
\hspace{2mm}
\subfloat[][$\vert\xi_{5}^{\operatorname{odd}}\rangle$]{\includegraphics[width=0.22\textwidth]{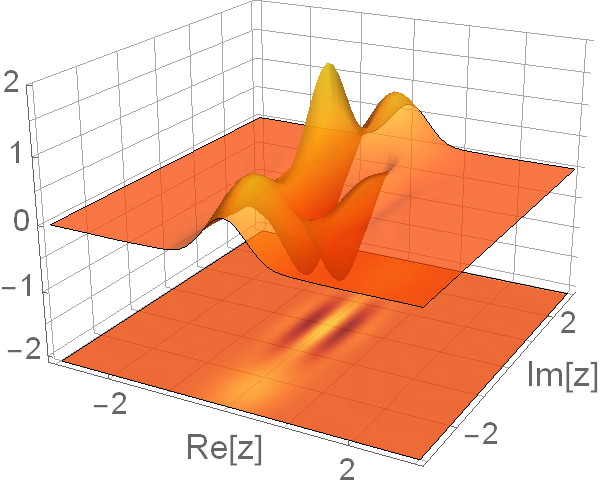}
\label{FWigf}}

\caption{\footnotesize Wigner distributions associated to the even-photon subtracted $\vert\xi_{2n}^{\operatorname{odd}}\rangle$ (first row) and odd-photon subtracted $\vert\xi_{2n+1}^{\operatorname{odd}}\rangle$ (second row) states generated out of the odd-photon squeezed states $\vert\xi^{odd}\rangle$. The latter are depicted in terms of the optical phase-space variable $z=\operatorname{Re} z +\mathrm{i}\operatorname{Im}z$, with the squeezing parameter fixed as $\vert\xi\vert=0.5$ in all figures.}
\label{FWig}
\end{figure}

Figure~\ref{FWig} depicts the behavior of the Wigner distribution $W(z)$~\cite{Ulf97} associated to the photon-subtracted states $\vert\xi_{n}^{\operatorname{odd}}\rangle$ (details to construct this distribution in simple form can be consulted in Appendix~A of~\cite{Zel18}). The $0$-photon subtracted case (Figure~\ref{FWiga}) reproduces qualitatively the Wigner distribution of the odd-photon squeezed state $\vert\xi^{\operatorname{odd}}\rangle$. The latter behaves like the distribution of the number state $\vert 1\rangle$ when it is squeezed along one of the optical phase-space variables, where the amount of squeezing is steered by $\vert\xi\vert$. In general, the $2n$-photon subtraction with $n>0$ induces the squeezing of the distribution by preserving the value of $\vert\xi\vert$ (Figures~\ref{FWigb}-\ref{FWigc}). After subtracting several photons the distribution exhibits a series of oscillations, which is a signature of nonclassical behavior. In contrast, the subtraction of $(2n+1)$- photons changes the behavior of the distribution drastically. This is expected as the final distribution becomes a combination of even number states, and the corresponding Wigner distribution behaves in a similar way to that of squeezed vacuum states $\vert\xi^{\operatorname{vac}}\rangle$. However, negative regions appear even when a few photons are subtracted, see Figures~\ref{FWigd}-\ref{FWigf}.

\subsection{Optical cat states}

Optical cat states are defined as the quantum superposition of two opposite-phase Glauber states \cite{Dod74}:
\bea
\vert z_E \rangle = \frac{e^{\frac{\vert z \vert^2}{2}}}{\lambda_E(z)} \left( \vert z \rangle + \vert - z \rangle \right)=
\frac{1}{\lambda_E(z)} \sum_{n=0}^{\infty} \frac{z^{2n}}{\sqrt{(2n)!}} \vert 2n \rangle, \quad \lambda_E (z) = \sqrt{ \cosh \vert z \vert^2 },
\label{cate}\\[1ex]
\vert z_O \rangle = \frac{e^{\frac{\vert z \vert^2}{2}}}{\lambda_O(z)} \left( \vert z \rangle + \vert - z \rangle \right)=
\frac{1}{\lambda_O (z)} \sum_{n=0}^{\infty} \frac{z^{2n+1}}{\sqrt{(2n+1)!}} \vert 2n +1\rangle, \quad \lambda_O (z)= \sqrt{ \sinh \vert z \vert^2 }.
\label{cato}
\eea
These nonclassical states \cite{Dod03} have been successfully created in laboratory by subtracting photons from the squeezed-vacuum state \cite{Our06,Daj97,Nee06,Wak07,Ger10,Nam10,Asa17}. 

Using  $\vert 0 \rangle_L = \vert - z \rangle$ and $\vert 1 \rangle_L = \vert z \rangle$ as logical qubits \cite{Obr09}, the optical cats are particularly useful for quantum information processing \cite{Our06}, where they are called {\em Schr\"odinger kittens} for $\vert z \vert \approx 1$.

The results in Section~\ref{vacio} apply immediately to analyze the related experimental data. As indicated above, the results in Section~\ref{oddsec} represent a secondary option. Note however that the optical cat states (\ref{cate}) and (\ref{cato}) are formally different from both, the squeezed-vacuum $\vert \xi^{\operatorname{vac}} \rangle$ and the odd-squeezed $\vert \xi^{\operatorname{odd}} \rangle$ states. We have already mentioned that the main reason for such dissimilarity is that states $\vert z_E \rangle$ and $\vert z_O \rangle$, together with the Glauber states $\vert z\rangle$, belong to the space of solutions associated with a first-order difference equation \cite{Zel21}. In turn, states $\vert \xi^{\operatorname{vac}} \rangle$ and $\vert \xi^{\operatorname{odd}} \rangle$ are independent solutions of a second-order difference equation \cite{Zel21}.

Within the photon subtraction scheme, the action of the beam splitter on the even cat $\vert z_E \rangle$ is easily calculated from Eq.~(\ref{parte5}).  Considering $\hat a \vert z \rangle = z \vert z \rangle$, one gets the entangled two-mode state
\be
\vert Z_E \rangle_B = \frac{1}{\sqrt{2}} \left[ 
\left\vert \left( \tfrac{iz}{\sqrt 2} \right)_{\!\! E} \right\rangle_a \otimes \left\vert \left( \tfrac{z}{\sqrt 2} \right)_{\!\! E} \right\rangle_b  
+ 
\left\vert \left( \tfrac{iz}{\sqrt 2}\right)_{\!\! O} \right\rangle_a \otimes \left\vert \left( \tfrac{z}{\sqrt 2} \right)_{\!\! O} \right\rangle_b
\right]
\label{bell1}
\ee
For the odd cat $\vert z_O \rangle$ we use Eq.~(\ref{parte6}), which yields
\be
\vert Z_O \rangle_B = \frac{1}{\sqrt{2}} \left[
\left\vert \left( \tfrac{iz}{\sqrt 2} \right)_{\!\! O} \right\rangle_a \otimes \left\vert \left( \tfrac{z}{\sqrt 2} \right)_{\!\! E} \right\rangle_b  
+ 
\left\vert \left( \tfrac{iz}{\sqrt 2}\right)_{\!\! E} \right\rangle_a \otimes \left\vert \left( \tfrac{z}{\sqrt 2} \right)_{\!\! O} \right\rangle_b
\right].
\label{bell2}
\ee
Considering the logical qubits $\vert 0 \rangle_L = \vert z_E \rangle$ and $\vert 1 \rangle_L = \vert z_O \rangle$, we see that the above results are in the Bell-basis representation
\be
\vert Z_E \rangle_B = \tfrac{1}{\sqrt{2}} \left[ \vert 00 \rangle_L + \vert 11 \rangle_L \right], \quad \vert Z_O \rangle_B = \tfrac{1}{\sqrt{2}} \left[ \vert 10 \rangle_L + \vert 01 \rangle_L \right],
\label{bell}
\ee
so they represent two maximally entangled quantum states of a two-qubit bipartite system. The construction of  Bell states as the result of passing optical cats by a beam splitter shows that the difference between the pairs $\vert Z_E \rangle$, $\vert Z_O \rangle$, and  $\vert \xi^{\operatorname{vac}} \rangle$, $\vert \xi^{\operatorname{odd}} \rangle$, is not merely formal. In the latter case entanglement is found between the elements of the superpositions defining the output modes, and not between the output superpositions by themselves.

In the language of quantum communication the results of Eq.~(\ref{bell}) read as follows: If Alice (mode $a$) and Bob (mode $b$) measure their qubit then both of them find a random result, either $\vert 0 \rangle_L$ or $\vert 1 \rangle_L$ with probability $\frac12$. Once Alice communicates her result to Bob (or vice versa), they find that their results are perfectly correlated, although their own outcomes seemed random. In the present case Alice does not require to `read' her qubit entirely. She needs to count the related number of photons only. In the even case, if she finds an even number of photons then Bob will find an even number of photons, with certainty. Similarly, if she finds an odd number of photons then Bob will read odd photons. A combined lecture of the number of photons is achieved in the odd case, since Alice will read an odd number of photons while Bob counts an even number, and vice versa.

\section{Conclusions}
\label{conclu}

We have studied how a two-mode entangled light beam can be produced by injecting superpositions of photon-number states into a lossless symmetric beam splitter. As the process changes the quantum statistics of the input fields, nonclassical correlations are stimulated between the output modes whenever the incident beam is nonclassical. We have considered the expansion of the output field in terms of the two-mode representation of $SU(2)$ coherent states, which encode nonclassical correlations between their modes. We derived a compact expression for the output field that is useful to construct photon subtracted versions of any input superposition of photon-number states. As applications, we derived the analytic form that links the detection of $n$-photons in the idler channel with the $n$-subtracted version of the squeezed-vacuum state in the signal channel. A second example provided the $n$-photon subtracted form of the odd-vacuum states, which represents a new class of nonclassical states. We have also shown that photon subtraction of classical fields do not produce nonclassical fields, and that injecting optical cats into a beam splitter produces entangled states in the Bell-basis representation.

\appendix
\section{Supplementary material}
\label{ApA}

\renewcommand{\thesection}{A-\arabic{section}}
\setcounter{section}{0}  

\renewcommand{\theequation}{A-\arabic{equation}}
\setcounter{equation}{0}  

The first basis elements (\ref{su2}) are as follows
\be
\begin{array}{l}
\vert 0, 0 \rangle_B = \vert 0,0 \rangle, \quad \vert 1, 0 \rangle_B = \frac{1}{\sqrt 2} \left (\vert 0,1 \rangle + i \vert 1,0 \rangle \right), \\[2ex]
\vert 2,0 \rangle_B = \frac12 \left( \vert 0,2 \rangle + i \sqrt 2 \vert 1,1 \rangle - \vert 2, 0 \rangle \right), \\[2ex]
 \vert3,0 \rangle_B= \frac{1}{2\sqrt 2} \left( \vert 0,3 \rangle + i\sqrt{3} \vert 1,2 \rangle - \sqrt{3} \vert 2,1 \rangle -i \vert 3,0 \rangle \right).
 \label{basis}
\end{array}
\ee
These vectors encode nonclassical correlations between modes $a$ and $b$. For instance, $\vert 1,0 \rangle_B$ in (\ref{basis}) is one of the elements in the Bell basis, which is prototypical to describe entanglement in bipartite qubit systems.

The two-mode state (\ref{state}) can be expressed as follows
\be
\vert \Psi  \rangle_B= \frac{1}{\lambda} \sum_{n,k=0}^{\infty} 
\alpha_{k+n} c_{n,k} \vert n \rangle_a \otimes \vert k \rangle_b,
\label{parte1}
\ee
where
\be
c_{n,k} =  \frac{1}{\sqrt{2^{n+k}}} \left( 
\begin{array}{c}
n+k\\n
\end{array}
\right)^{1/2} e^{i \frac{\pi}{2} n}.
\label{arreglo}
\ee
Then
\be
\vert \Psi  \rangle_B= \frac{1}{\lambda} \sum_{n,k =0}^{\infty} 
\tfrac{\sqrt{(k+n)!}}{\sqrt{n!} \sqrt{k!}} e^{i\frac{\pi}{2}n} \beta_{k+n}
\vert n \rangle_a 
\otimes \vert k \rangle_b, \quad \beta_n = \tfrac{\alpha_n}{\sqrt{2^n}}.
\label{state2}
\ee
Using Eq.~(\ref{sus1}) we have either 
\[
\vert \Psi  \rangle_B= \frac{1}{\lambda} \sum_{n=0}^{\infty} 
\frac{e^{i\frac{\pi}{2}n}}{\sqrt{n!}} \lambda_n (\beta) \vert n \rangle_a 
\otimes \vert \psi_n(\beta) \rangle_b,
\]
or 
\[
\vert \Psi \rangle_B= \frac{1}{\lambda(\alpha)} \sum_{k =0}^{\infty} 
\frac{e^{-i\frac{\pi}{2}k}}{\sqrt{k!}} \lambda_k(\beta) \vert \psi_k (\beta) \rangle_a \otimes \vert k \rangle_b.
\]
These two expressions are equivalent.

For input states expressed as linear superpositions consisting entirely of even-photon states
\[
\vert \psi_E \rangle = \frac{1}{\lambda_E}\sum_{n=0}^{\infty} \alpha_{2n} \vert 2n \rangle 
\]
we have
\be
\begin{array}{l}
\vert \Psi_E \rangle_B = \displaystyle\frac{1}{\lambda_E} \sum_{n,k=0}^{\infty}  
\left[ 
\alpha_{2(k+n)} c_{2n,2k} \vert 2n \rangle_a \otimes  \vert 2k \rangle_b  
\right. \\[3ex]
\hskip5cm
\left.
+ \alpha_{2(k+n+1)} c_{2n+1,2k+1}  \vert 2n +1 \rangle_a \otimes  \vert 2k+1 \rangle_b
\right],
 \end{array}
\label{parte3}
\ee
Using (\ref{arreglo}) and (\ref{sus1}) yields
\[
\begin{array}{l}
\vert \Psi_E \rangle_B =\displaystyle\frac{1}{\lambda_E} \sum_{n=0}^{\infty}  
\left[ \tfrac{e^{i\pi n}}{\sqrt{(2n)!}} \lambda^E_{2n} (\beta) \vert 2n \rangle_a \otimes \vert \psi^E_{2n} (\beta) \rangle_b 
\right. \\[3ex]
\hskip5cm
\left.
+ \displaystyle \tfrac{i e^{i\pi n}}{\sqrt{(2n+1)!}} \lambda^E_{2n+1} (\beta) \vert 2n+1 \rangle_a \otimes \vert \psi^E_{2n+1} (\beta) \rangle_b 
\right],
\end{array}
\]
where
\be
\vert \psi_{2n}^E (\alpha) \rangle_b= \frac{1}{\lambda_{2n}^E  (\alpha)} \sum_{k=0}^{\infty} \sqrt{ \tfrac{(2 k+2n)!}{(2k)!}} \alpha_{2(k+n)} \vert 2k \rangle_b,
\label{sus2even}
\ee
and
\be
\vert \psi_{2n+1}^E (\alpha) \rangle_b= \frac{1}{\lambda_{2n+1}^O  (\alpha)} \sum_{k=0}^{\infty} \sqrt{ \tfrac{(2 k+2n+2)!}{(2k+1)!}} \alpha_{2(k+n+1)} \vert 2k +1\rangle_b.
\label{sus2odd}
\ee

The case of superpositions including only odd-photon states 
\[
\vert \psi_O \rangle = \frac{1}{\lambda_O}\sum_{n=0}^{\infty} \alpha_{2n+1} \vert 2n+1 \rangle
\]
gives the result
\be
\begin{array}{l}
\vert \Psi_O \rangle_B = \displaystyle\frac{1}{\lambda_O} \sum_{n,k=0}^{\infty}
\left[ 
\alpha_{2(k+n)+1} c_{2n,2k+1} \vert 2n \rangle_a \otimes \vert 2k + 1 \rangle_b
\right. \\[3ex]
\hskip5cm
\left.
+ \alpha_{2(k+n)+1} c_{2n+1,2k} \vert 2n +1 \rangle_a \otimes \vert 2k \rangle_b
\right],
\end{array}
\label{parte4}
\ee
which is simplified as follows
\[
\begin{array}{l}
\vert \Psi_O \rangle_B = 
\displaystyle\frac{1}{\lambda_O} \sum_{n=0}^{\infty}
\left[
\tfrac{e^{i \pi n}}{\sqrt{(2n)!}} \lambda_{2n}^O( \beta)
\vert 2n \rangle_a \otimes \vert \psi_{2n}^O( \beta)  \rangle_b
\right. \\[3ex]
\hskip5cm
\left.
+  \displaystyle \tfrac{i e^{i \pi n}}{\sqrt{(2n+1)!}} \lambda_{2n+1}^O( \beta)
\vert 2n +1 \rangle_a \otimes \vert \psi_{2n+1}^O( \beta)  \rangle_b
\right],
\end{array}
\]
where
\be
\vert \psi_{2n}^O (\alpha) \rangle_b= \frac{1}{\lambda_{2n}^O  (\alpha)} \sum_{k=0}^{\infty} \sqrt{ \tfrac{(2 k+2n+1)!}{(2k+1)!}} \alpha_{2(k+n)+1} \vert 2k+1 \rangle_b,
\label{sus3even}
\ee
and
\be
\vert \psi_{2n+1}^O (\alpha) \rangle_b= \frac{1}{\lambda_{2n+1}^O  (\alpha)} \sum_{k=0}^{\infty} \sqrt{ \tfrac{(2 k+2n+1)!}{(2k)!}} \alpha_{2(k+n)+1} \vert 2k \rangle_b.
\label{sus3odd}
\ee

\section*{Acknowledgments}

This research has been funded by Consejo Nacional de Ciencia y Tecnolog\'ia (CONACyT), Mexico, Grant Numbers A1-S-24569 and CF19-304307.

K. Zelaya acknowledges the support from the project ``Physicist on the move II'' (KINE\'O II), Czech Republic, Grant No. CZ.02.2.69/0.0/0.0/18053/0017163.


\end{document}